\documentclass[twocolumn,preprintnumbers,prb]{revtex4}

\usepackage{graphicx}% Include figure files
\usepackage{bm}% bold math
\usepackage{amsmath}
\usepackage{amssymb}
\begin{document}

%\preprint{PRL/001-BSD}

\title{Nanoscale characterization of bismuth telluride epitaxic layers by advanced X-ray analysis}

\author{S\'ergio L. Morelh\~ao}
\email[corresponding author: ]{morelhao@if.usp.br}
\affiliation{Instituto de F\'isica, Universidade de S\~ao Paulo, S\~ao Paulo, SP, Brazil}
\author{Celso I. Fornari}
\affiliation{Instituto Nacional de Pesquisas Espaciais, LAS, S\~ao Jos\'e dos Campos, SP, Brazil}
\author{Paulo H. O. Rappl}
\affiliation{Instituto Nacional de Pesquisas Espaciais, LAS, S\~ao Jos\'e dos Campos, SP, Brazil}
\author{Eduardo Abramof}
\affiliation{Instituto Nacional de Pesquisas Espaciais, LAS, S\~ao Jos\'e dos Campos, SP, Brazil}

\date{\today}

\begin{abstract}
Topological insulator surface properties are strongly correlated to structural properties, requiring high-resolution techniques capable of probing both surface and bulk structures at once. In this work, high flux of synchrotron source, recursive equations for fast X-ray dynamical diffraction simulation, and genetic algorithm for data fitting are combined to reveal the detailed structure of bismuth telluride epitaxic films with thickness ranging from 8 to 168 nm. It includes stacking sequences, thickness and composition of layers in model structures, interface coherence, surface termination and morphology. These results are in agreement with the surface morphology determined by atomic force microscopy. Moreover, by using X-ray data from zero noise area detector to construct three-dimensional reciprocal space maps, insights into the nanostructure of domains and stacking faults in Bi$_2$Te$_3$ films are given.
\end{abstract}

%\pacs{Valid PACS appear here}

\maketitle

\section{Introduction}

Bismuth chalcogenide compounds have recently attracted great attention due to their properties as a three-dimensional topological insulator. This new class of materials is insulating in the bulk and exhibits gapless metallic surface states with linear energy-momentum dispersion shaped like a Dirac cone. Due to the strong spin-orbit coupling, these conducting surface states have electron momentum locked to the spin orientation and are protected from scattering mechanisms by time reversal symmetry. Hence, in the absence of external magnetic fields, high-mobility spin polarized surface currents can be produced, offering possibilities to new applications in spintronics \cite{has10,and13}. Particularly in Bi$_2$Te$_3$ material, topological features have been theoretically predicted \cite{zha09} and the experimental observation of the metallic surface states, consisting of a single Dirac cone at the $\Gamma$ point, has been demonstrated by angle resolved photoelectron spectroscopy measurements in bulk crystals \cite{che09,hsi09a,hsi09b}. Insulating bulk samples can be obtained by counter doping with Sn since it moves the Fermi level to inside the band gap \cite{che09}. Furthermore, surface states are also present in high-quality epitaxial films grown by molecular beam epitaxy (MBE) \cite{wan11,hoe14}. By controlling the MBE growth parameters, thin films with intrinsic conduction through only topological surface states can be obtained, i.e., films where the Fermi level is crossing only the V-shaped Dirac cone \cite{wan11,hoe14}. The metallic states are susceptible to surface processes and, consequently, to surface structures  \cite{ban00,par10,yas13,kuz15}. In this sense, it is desirable to have a fine structural characterization technique able to determine the surface chemical termination, as well as the nanostructure of domains in topological insulator thin films.

\subsection{General problems in nanostructure characterization}

Nanoscale manufacturing for the modern semiconductor and optical industries makes broad  usage of thin crystalline films and multilayers. Quality control and development of new electronic and optical devices demand suitable methods for characterization of these structures. Simulations of X-ray scattering and diffraction are well-established procedure for structural analysis at nanometer and subnanometer length scales of layered materials, ranging from amorphous films to  crystalline ones such as epitaxial layers on single-crystal substrates. Higher are the ordering in stacking sequences of the atomic layers, the more pronounced are the diffracted intensities at higher angles allowing more refined model structures. X-ray theories are well comfortable at the limiting cases, either amorphous films or perfect periodic layer sequences, i.e. crystalline films. However, in developing new materials and processing technologies, layered materials with random layer sequences of large \textit{d}-spacing can often be found. Combined with the very high dynamical range of advanced X-ray sources and detectors, this kind of material represent a challenging in theoretical approach for X-ray diffraction simulation \cite{ale08}.

Large \textit{d}-spacing implies that diffraction peaks are relatively close to each other, compromising theoretical approaches that treat them separately. On the other hand, low order or lack of perfect periodicity produces a systematic degradation of intensity signal in diffraction peaks at higher angles. The simple kinematic approach that neglects refraction and rescattering events of diffracted photons can be good  enough for weak intensity reflections and even regions in between peaks. But, strong peaks at lower angles as well as diffraction peaks from single-crystal substrates may require an approach accounting for effects of refraction and rescattering. A similar situation is found in the investigation of surface structures by scanning of crystal truncation rods \cite{ac94,fei89,kag07,vh08,coe13}. At the moment, these effects are accounted for in dynamical diffraction theories suitable for very crystalline materials. Although long-range scans in reciprocal space, overlapping of diffraction peaks, and even strain in the crystal lattice have been treated within the scope of dynamical diffraction \cite{ale08,kag07,gro91}, they are distinct approaches and potential users are discouraged by the mathematical complexity that has to be understood to adapt these approaches to each particular system under investigation. In this scenario, it is of great importance to have an approach that is as simple as the kinematic one, able to account for refraction, absorption, and rescattering in any kind of layered material, and also easily implemented in computer routines for fast simulation of diffraction experiments.

Methods to implement nanostructured models for simulation of X-ray diffraction in layered materials are described here and applied to study the case of bismuth telluride epitaxial films grown on barium fluoride (111) substrates. A few of such models have already been successfully applied to quantify structural information of our bismuth telluride epitaxial films\cite{for16,cif16}. Long-range scans from model structures are calculated recursively and compared to the actual X-ray diffraction data. Fast data calculation allow best-fit procedures for determining the most probable stacking sequence of atomic monolayers along the epitaxial films, as well as the fraction of surface areas covered with different layer thicknesses.

\section{Model structures}

\begin{figure}
  \includegraphics[width=2.7in]{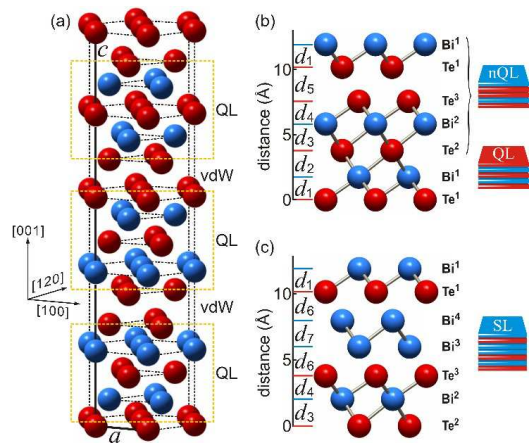}\\
  \caption{(a) Hexagonal crystalline structure of Bi2Te3. The unit cell is formed by stacking of three quintuple layers (QL) along [001] direction. The van der Waals (vdW) coupling between adjacent QLs through Te atoms is much weaker than that between Te and Bi inside the QL. Atomic inter-layer distances in (b)  quintuple layer (QL), non-conventional QL (nQL), and (c) septuple layer (SL) sets of monolayers used here to simulate X-ray diffraction in epitaxial films.}
\label{fig:nQLSL}
\end{figure}

Bi$_2$Te$_3$ crystallizes in a tetradymite-type structure shown in Fig.~\ref{fig:nQLSL}a. The unit cell of this hexagonal lattice is described by stacking of three Te$^1$:Bi$^1$:Te$^2$:Bi$^2$:Te$^3$ quintuple layers (QLs) along the $c$ direction, with a lattice parameter $c$ = 30.474\,\AA\, for the bulk material. Inside the QL, the neighboring Bi:Te atoms are ionic bonded and the adjacent QLs are van der Waals coupled to each other. The van der Waals gaps give to the structure an anisotropic character similar to other layered materials. In-plane lattice constant of the BaF$_2$ (111) surface $a_S^{(111)}=a_S/\sqrt{2}= 4.384$\,\AA\, is almost equal to the hexagonal lattice constant $a$ = 4.382\,\AA\, of Bi$_2$Te$_3$. This small lattice mismatch of only 0.04\% between both materials makes BaF$_2$ (111) a suitable substrate for the epitaxy of bismuth telluride. In this epitaxy, the Bi$_2$Te$_3$ (001) hexagonal planes are parallel to the cubic BaF$_2$ (111) planes and only the (00L) Bragg peaks with L=3$n$, where $n$ is a positive integer, are allowed. Epitaxial films of Bi$_2$Te$_{3-\delta}$ with a deficit $\delta$ of Te in the range $0<\delta<1.5$ have Bi:Bi bilayers (BLs) inserted in between QLs of the Bi$_2$Te$_3$ phase. The insertion of BLs during epitaxial growth does not occur uniformly, but as a statistical distribution where the grown films can be considered as a random one-dimensional Bi$_x$Te$_y$ alloy rather than an ordered homologous (Bi$_2$)$_{\rm M}$(Bi$_2$Te$_3$)$_{\rm N}$ structure with deficit $\delta$ = 3M/(M+N) \cite{for16,ste14}.

To simulate X-ray diffraction in such films, we consider two base sets of atomic monolayers (MLs) depicted in Figs.~\ref{fig:nQLSL}b and \ref{fig:nQLSL}c. One set is a non-conventional QL (nQL), Te$^2$:Bi$^2$:Te$^3$::Te$^1$:Bi$^1$ in Fig.~\ref{fig:nQLSL}b, in which the Te$^3$::Te$^1$ van der Waals gap is at the middle, and another set that is a septuple layer (SL) with the Bi$^3$:Bi$^4$ BL inserted in the van der Waals gap, Fig.~\ref{fig:nQLSL}c. Model structures starting and ending with conventional QLs are obtained by adding the required Te$^1$:Bi$^1$ and Te$^2$:Bi$^2$:Te$^3$ MLs to complete the QLs at the bottom and top of the nQL/SL stacking sequences. For instance, the sequence
\begin{equation}\label{eq:sequence}
{\rm Te}^1:{\rm Bi}^1:\left\{{\rm nQL}_{J-1}:{\rm SL}\right\}_K:{\rm nQL}_{J-1}:{\rm Te}^2:{\rm Bi}^2:{\rm Te}^3
\end{equation}
gives rise to the ordered $({\rm Bi}_2)_K ({\rm Bi}_2{\rm Te}_3)_{J(K+1)}$ structure with composition $\delta=3K/[J(K+1)+K]$. In thick structures, $K\gg1$, the maximum Te deficit that can be obtained with this sequence is for $J=1$ where $\delta=3/2$, corresponding to the $({\rm Bi}_2)_1 ({\rm Bi}_2{\rm Te}_3)_1$ structure, i.e. the Bi$_4$Te$_3$ phase. By using the sequence in Eq.~(\ref{eq:sequence}) with $K=0$, a pure Bi$_2$Te$_3$ structure with $J$ complete QLs is obtained. Other models with different surface terminations, such as Bi$^3$:Bi$^4$ and Bi$^3$:Bi$^4$:Te$^1$ that have been observed in bulk crystals \cite{coe13}, can also be considered when fitting the experimental data.

One advantage of using the nQL and SL sets of MLs is that in any random stacking sequence there will be at least one QL sandwiched in between two consecutive BLs, i.e. 4 adjacent MLs of Bi never occur in our structure models of the films. Another advantage of using nQLs and SLs to describe the film structure is that the inter-layer distance to stack any of these sets upon each other is $d_2$, corresponding to the distance between Bi$^1$:Te$^2$ since both sets have the ML Te$^2$ at the bottom and the Bi$^1$ at the top, Figs.~\ref{fig:nQLSL}b and \ref{fig:nQLSL}c. This fact is of practical importance when computing X-ray diffraction in model structures.

The inter-layer distance $d_2$ is nearly constant with respect to the film composition, as well as the other distances indicated in Fig.~\ref{fig:nQLSL}b, $d_1=d_4= 1.746$\,\AA, $d_2 = d_3 = 2.035$\,\AA, and $d_5 = 2.613$\,\AA\, at room temperature. The small in-plane strain developed as a function of composition is accounted for in $d_6 = 2.152+0.075\,\delta$\,\AA\, and $d_7 = 2.003-0.075\,\delta$\,\AA, Fig.~\ref{fig:nQLSL}c, which is a solution for the variation of mean inter-layer atomic distance $\langle d \rangle(\delta)=2.035-0.025\,\delta$\,\AA\, reported in a series of Bi$_2$Te$_{3-\delta}$ epitaxial films deposited on BaF$_2$ (111) substrate by molecular beam epitaxy \cite{for16,ste14}. Other solutions accounting for variations in $d_1$ and $d_5$ are also possible, as discussed in \S\ref{ap:ipdist}.

\section{Recursive series}

To account for the variation of the X-ray wave as it crosses a single atomic ML, as well as the multiple reflections that occur in between any two adjacent MLs in layered materials, the following approach is used \cite{slm16}. If $r_A$ and $t_A$ are the reflection and transmission coefficients of amplitude for a monochromatic X-ray wave reaching a given ML, labeled A, the coefficients for a double layer formed by A and B types of MLs will be
\begin{eqnarray}\label{eq:receqs}
  r_{AB} &=& r_A+r_B\dfrac{t_A t_A\,e^{2 i \varphi}}{1 - \bar{r}_A r_B\,e^{2 i \varphi}}\;, \nonumber \\
  \bar{r}_{AB} &=& \bar{r}_B+\bar{r}_A\dfrac{t_B t_B\,e^{2 i \varphi}}{1 - \bar{r}_A r_B\,e^{2 i \varphi}}\;,\quad{\rm and} \nonumber \\
  t_{AB} &=& \dfrac{t_A t_B\,e^{i \varphi}}{1 - \bar{r}_A r_B\,e^{2i \varphi}}\;.
\end{eqnarray}
$\varphi=-\frac{1}{2}Qd$ is the phase delay every time the X-ray wave of wavelength $\lambda$ crosses the inter-layer distance $d$ between the MLs, and $Q=(4\pi/\lambda)\sin\theta$ is the modulus of the scattering vector perpendicular to the MLs for an incidence angle $\theta$. Reflection coefficients of layered structures having more than one ML can be different when the X-ray impinges from the top, coefficient $r_{AB}$, or from the bottom, coefficient $\bar{r}_{AB}$. But, for a single atomic ML they are identical, i.e. $r_A=\bar{r}_A$ and $r_B=\bar{r}_B$.

\begin{figure}
\includegraphics[width=2.7in]{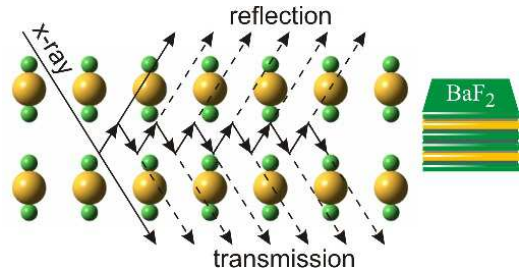}\\
\caption{Multiple rescattering of X-ray photons between atomic layers in BaF$_2$ (111).}
\label{fig:BaF2stack}
\end{figure}

Accounting for multiple rescattering in between the MLs produces a geometrical series $\sum_{n=0}^\infty x^n = 1/(1-x)$ of ratio $x=\bar{r}_A r_B\,e^{2 i \varphi}$, as schematized in Fig.~\ref{fig:BaF2stack}. By neglecting this series, i.e., by taking $\bar{r}_A r_B=0$ in the denominators of Eqs.~\ref{eq:receqs}, we basically end up with the kinematic approach, although with corrections for refraction and absorption implicit in the reflection and transmission coefficients of each atomic ML, see \S\ref{ap:rescatt} for a more detailed analysis of the effects of rescattering. In the case of a general ML X, containing more than one atomic species, the coefficients are \cite{cgd14,ac94}  $r_X=-i\Gamma\sum_a\eta_a f_a(Q,E)$ and $t_X=1+i\Gamma\sum_a\eta_a f_a(0,E)$ where $\eta_a$ is the area density of atoms $a$ in the ML plane and $f_a(Q,E)=f_a^0(Q)+f_a^{\prime}(E)+if_a^{\prime\prime}(E)$ are their atomic scattering factors with resonant amplitudes for X-ray photons of energy $E$, see \S\ref{ap:scattmls}. The parameter $\Gamma=r_e\lambda C/\sin\theta$ arises from the scattering and photoelectric absorption cross sections, and it is very small due to the low value of electron radius $r_e = 2.818\times10^{-5}$\,\AA. The polarization term is $C$, and $\sin\theta$ takes into account area variation of the beam footprint at the sample surface. The real and imaginary terms of $t_X$ provide absorption and refraction corrections, respectively, while the resonant amplitudes in $r_X$ imply that $r_{AB}\neq\bar{r}_{AB}$ in structures with more than one type of MLs \cite{slm16}.

\begin{figure*}
 \includegraphics[width=5.4in]{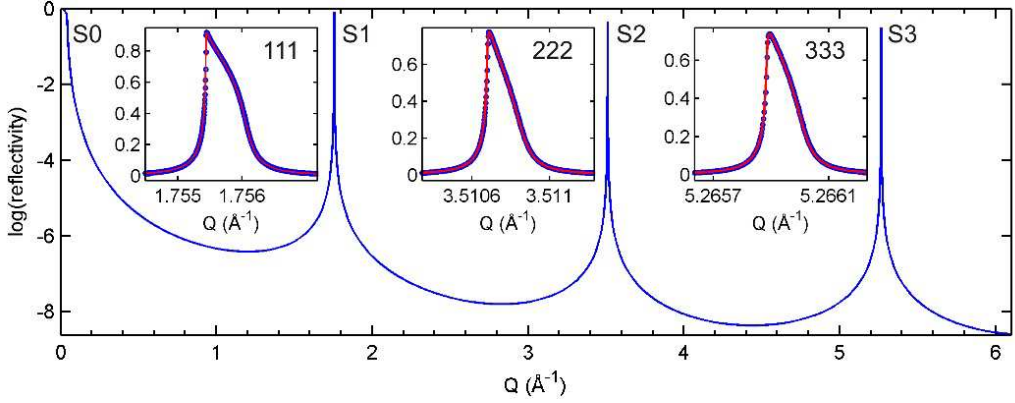}\\
 \caption{X-ray reflectivity in a 93.8\,$\mu$m thick BaF2(111) substrate calculated with recursive equations, Eqs.~\ref{eq:receqs}. Narrow $Q$-scans around each diffraction peak by using either the recursive equations (blue circles) or dynamical theory (red lines) are shown in the insets. Labels S1, S2, and S3 correspond to 111, 222, and 333 substrate Bragg peaks, respectively, while S0 is the total refraction at grazing incidence.}
 \label{fig:BaF2Qscan}
\end{figure*}

To add a third ML of another type, e.g. ML C, at the bottom of the A:B double layer, Eqs.~\ref{eq:receqs} can be used recursively by replacing the coefficients of ML A with the double layer coefficients, and the coefficients of ML B with those of ML C. The inter-layer distance $d$ used in this case is the distance between the B and C MLs. It results in the coefficients for the triple layer A:B:C. Note that in perfect periodic structures where a base sequence of a few MLs is repeated many times, the coefficients for thick materials are computed very fast since application of Eqs.~\ref{eq:receqs} $N$ times result in $2^N$ repetitions of the same base set of MLs. Comparison of this recursive procedure with the well known dynamical theory of X-ray diffraction \cite{bat64,wec97,aut06} is shown in Fig.~\ref{fig:BaF2Qscan} for a 93.8\,$\mu$m ($N$=18) thick BaF$_2$ (111) substrate where S1, S2, and S3 stand for the 111, 222, and 333 Bragg peaks, respectively. Along the [111] direction, the base set of MLs is the triple layer F:Ba:F that repeats with periodicity $a_S/\sqrt{3}$ where $a_S = 6.2001$\,\AA\, is the BaF$_2$ cubic lattice parameter.

There are a few fundamental differences in the approach summarized by Eqs.~(\ref{eq:receqs}) regarding the most similar ones aimed to simulate the entire reflectivity curve from grazing to normal angles of incidence \cite{ac94,vh08}. Basically, here dynamical diffraction corrections due to photon rescattering have been solved in between any two atomic layers, independently of the stacking sequence, i.e., of ordering and neighboring layers. This solution holds likewise either in periodic or completely random sequences. No concern in accounting individual Bragg reflections is required. They appear naturally as a consequence of order or periodicity. As formalized in Eqs.~(\ref{eq:receqs}), this approach is limited to the specular reflection geometry. It gives up generalism in favor of simplicity, a fact that can be of practical importance in studying many systems of layered materials.

\subsection{Simulation of X-ray diffraction in epitaxial films}

Diffracted waves from bulk substrates as well as from films are provided by Eqs.~\ref{eq:receqs}, which are also applied to combine both waves. By taking $r_F$ and $t_F$ as reflection and transmission coefficients of the entire film structure, and $r_S$ as reflection coefficient of the substrate before the film, the film/substrate system will have a final reflectivity given by
\begin{equation}\label{eq:R}
    R=|A|^2+|B|^2 + \chi(AB^*+A^*B)
\end{equation}
where $A=r_F$, and $B=r_S t_F t_F e^{2 i\varphi}/(1-\bar{r}_F r_S e^{2 i\varphi})$, and $\varphi=-\frac{1}{2}Qd_i$. The value of $\chi = 1$ is for sharp interfaces where the film/substrate distance $d_i$ is constant over the sample area, while $\chi = 0$ is for samples with rough interfaces where $d_i$ fluctuates in a magnitude larger than $4\pi/Q$. The parameter $\chi$ gauges the average interference between diffracted waves from substrate and film lattices. Small surface roughness can be accounted for in standard Debye-Waller factors, multiplying the film and substrate reflection coefficients.

\begin{figure*}
  \includegraphics[width=5.4in]{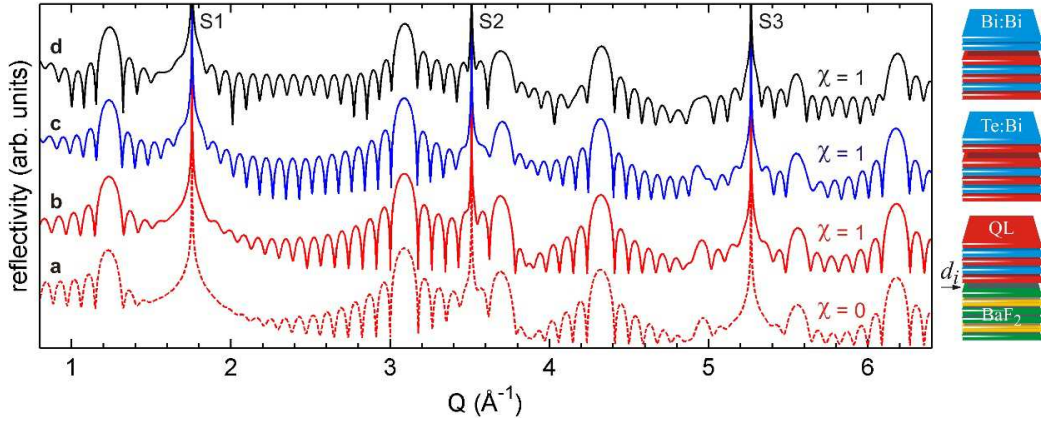}\\
  \caption{Simulation of X-ray diffraction in films with 7 QLs on BaF$_2$ (111) substrate and interface distance $d_i = 1.87$\,\AA. Curves \textbf{a,b}: comparison of considering $\chi = 0$ or $\chi = 1$ in Eq.~\ref{eq:R} for surface termination in QL. Curves \textbf{b-d}: comparison of surface termination in QL, QL + Te:Bi, and QL+Bi:Bi for $\chi = 1$.}
\label{fig:7QLs}
\end{figure*}

Fig.~\ref{fig:7QLs} shows a few examples on how X-ray diffraction curves of bismuth telluride thin films grown on BaF$_2$ (111) are susceptible to the interface quality and surface termination. The influence of roughness on the interface between substrate and layer with a fixed interface distance $d_i$ is illustrated by curves \textbf{a} and \textbf{b}, Fig.~\ref{fig:7QLs}. They show the simulation for a 7 QLs thick Bi$_2$Te$_3$ film terminated in a perfect QL with a completely rough interface ($\chi = 0$) and a sharp interface ($\chi = 1$), respectively. Note that the main differences between both curves remain in the neighborhood of the substrate peaks. Curves \textbf{b-d} exemplify the effect of surface termination on the simulation of X-ray diffraction for a 7 QLs thick film with $\chi = 1$. In this case, the surface termination modifies the intensity and the shape of the layer diffraction peaks, the interference fringes pattern between these peaks, and also the vicinity of the substrate peaks. Henceforward, only substrate termination in complete F:Ba:F set of MLs is considered since it is the most probable one on cleaved (111) surfaces \cite{shi06}.

\begin{figure*}
  \includegraphics[width=5.4in]{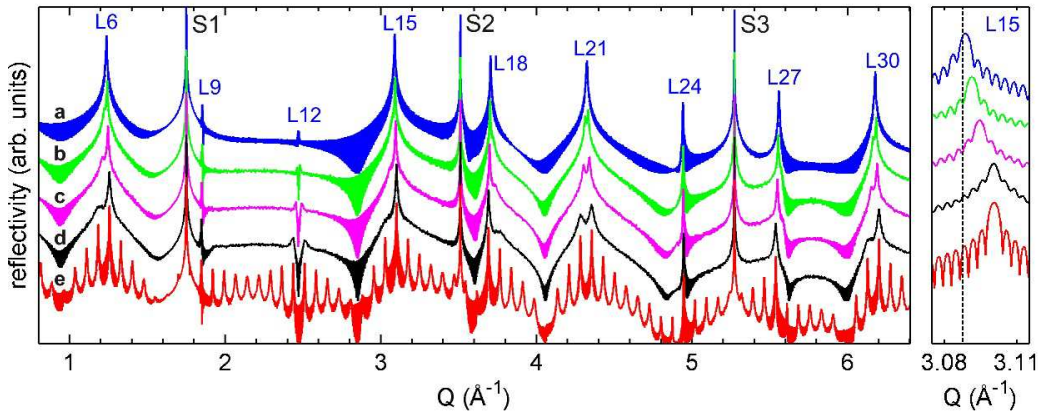}\\
  \caption{Simulation of X-ray diffraction in Bi$_2$Te$_{3-\delta}$ films on BaF$_2$ (111) substrate. All films have 168 QLs with a few BLs of bismuth randomly distributed along the film thickness: in curves \textbf{a-d}, 1 ($\delta$ = 0.018), 5 ($\delta$ = 0.087), 10 ($\delta$ = 0.168), and 20 ($\delta$ = 0.319) BLs, respectively. In curve \textbf{e}, a highly ordered structure with 1 BL for each 8 QLs is considered ($\delta$ = 0.319). Peak labels L refer to those of the pure Bi$_2$Te$_3$ structure and S to the substrate ones. L15 peak position (dashed line) is indicated in zoomed view at left.}
\label{fig:simQscanrand}
\end{figure*}

\begin{figure}
  \includegraphics[width=2.7in]{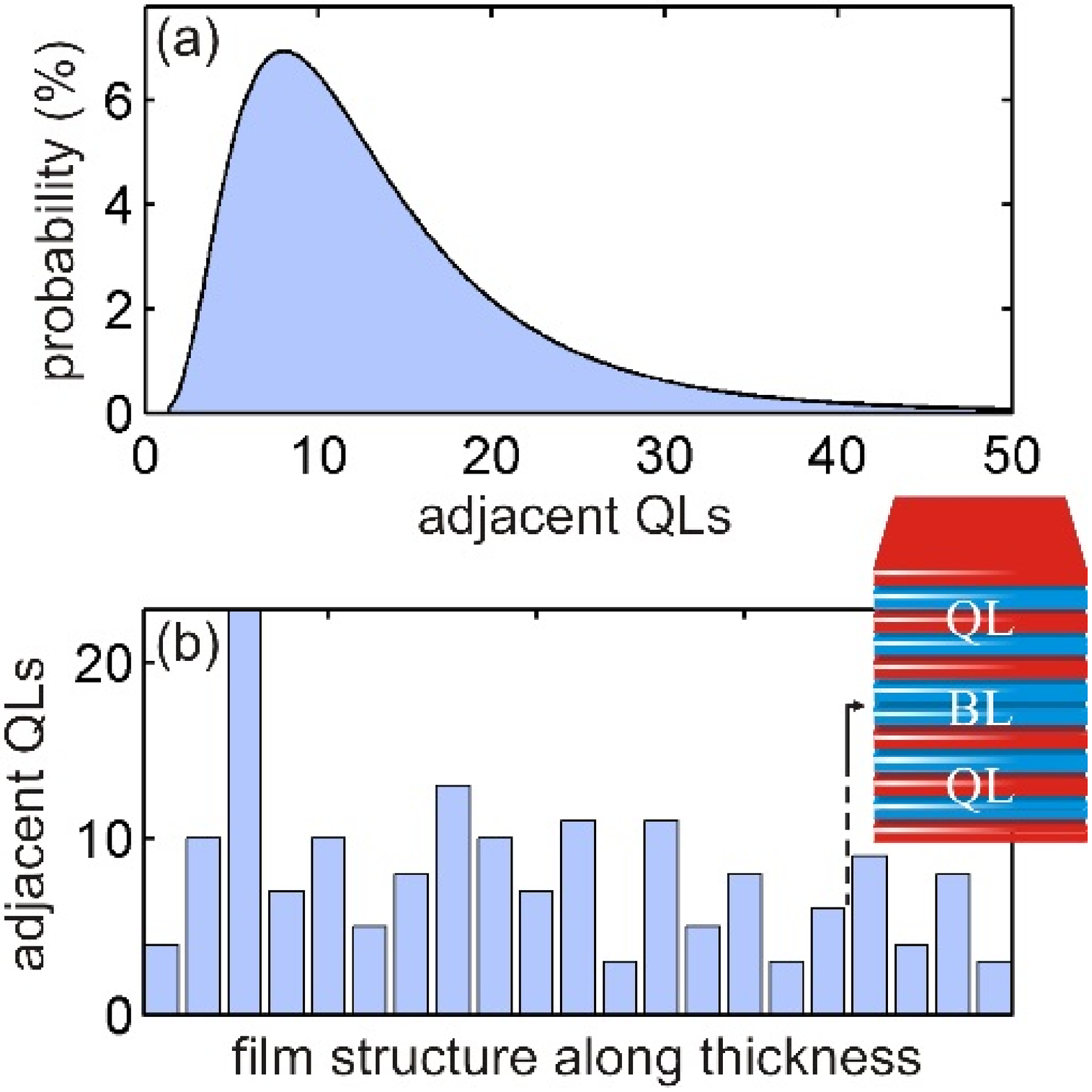}\\
  \caption{(a) Probability function for the number of adjacent QLs without a BL of bismuth. (b) Example of film structure according to this probability function, containing a total of 168 QLs (sum of bar heights) and 20 BLs (space in between bars). Each bar stands for the number of adjacent QLs without a BL, average of 1 BL for each 8 QLs, $\delta$ = 0.319.}\label{fig:QLdistrib}
\end{figure}

Theoretical investigation of effects on X-ray diffraction curves of gradually adding BLs of bismuth in the film structures is shown in Fig.~\ref{fig:simQscanrand}. Each curve is an average of two hundred curves computed with the BLs distributed along film thickness according to a log-normal probability function such as the one in Fig.~\ref{fig:QLdistrib}, see also \S\ref{ap:dfitt}. Well ordered structures are obtained by narrowing the probability function, which lead to similar stacking sequences to that described in Eq.~(\ref{eq:sequence}). For instance, the simulated curve \textbf{e} in Fig.~\ref{fig:simQscanrand} stands for a structure given by $K=20$ and $J=8$ in Eq.~(\ref{eq:sequence}). Besides the shifting of the L15 peak of the Bi$_2$Te$_3$ structure according to $2\pi/\langle d \rangle(\delta)$ \cite{ste14}, splitting $\Delta Q = 0.23\,\delta\,\textrm{\AA}^{-1}$ of the L21 peak is another effect that can be useful to estimate film composition \cite{for16}.

\section{Experimental}

Bismuth telluride films were grown on freshly cleaved (111) BaF$_2$ substrates in a Riber 32P molecular beam epitaxial (MBE) system using an effusion cell charged with nominal stoichiometric Bi$_2$Te$_3$ solid source and two additional Te cells to offer the extra Te flux. The substrate temperature and the Te to Bi$_2$Te$_3$ beam flux ratio were varied from 220$^{\circ}$C to 300$^{\circ}$C and from 0 to 2, respectively. These MBE growth conditions give rise to Bi$_2$Te$_{3-\delta}$ epitaxial films with Te deficit $\delta$ ranging from 0 to 0.3. Details about the MBE growth and structural characterization are published elsewhere \cite{for16}. Atomic force microscopy (AFM) images of the bismuth telluride films were taken ex situ in a Veeco Multimode V Scanning Probe Microscope using a silicon nitride probe in tapping mode.

X-ray diffraction measurements were made at the XRD2 beamline of the Brazilian Synchrotron Light Laboratory (LNLS). The beam was vertically focused with a bent Rh-coated mirror, which also filtered higher-order harmonics. The beam energy was tuned to 8004\,eV (1.549038\,\AA) using a double-bounce (111) Si monochromator, placed after the Rh mirror. The beam was focused at the sample position in a spot of 0.6\,mm (vertical) $\times$ 2\,mm (axial), with a flux of the order of $10^{10}$ photons/mm$^2$/s. The sample was mounted onto the Eulerian cradle of a Huber 4+2 circle diffractometer, with vertical scattering plane. X-ray diffraction data were collected at $\sigma$-polarization (vertical scattering plane) by a Pilatus 100 K area detector (pixel size of 172\,$\mu$m), used as a point detector with vertical and axial acceptances of about 0.05$^{\circ}$ and 0.1$^{\circ}$, respectively. Sample-detector distance was set to 910\,mm and air absorption minimized by using evacuated fly tubes.

\section{Results and discussions}

\begin{figure*}
  \includegraphics[width=5.4in]{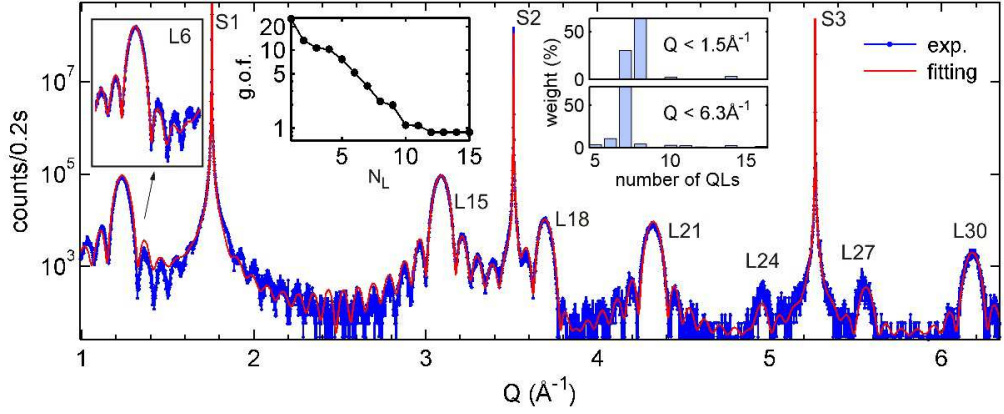}\\
  \caption{Experimental and fitted $Q$-scans along crystal truncation rods of 00L reflections in Bi$_2$Te$_3$ film on BaF$_2$ (111), sample S15058. Insets (from right to left): model structures in number of QLs covering the sample area (weight); goodness of fit (g.o.f.) value as function of the number $N_L$ of layers in the model structures; and best fit for the region $Q < 1.5\,\textrm{\AA}^{-1}$.}
\label{fig:Qscan15058}
\end{figure*}

\begin{figure*}
  \includegraphics[width=5.4in]{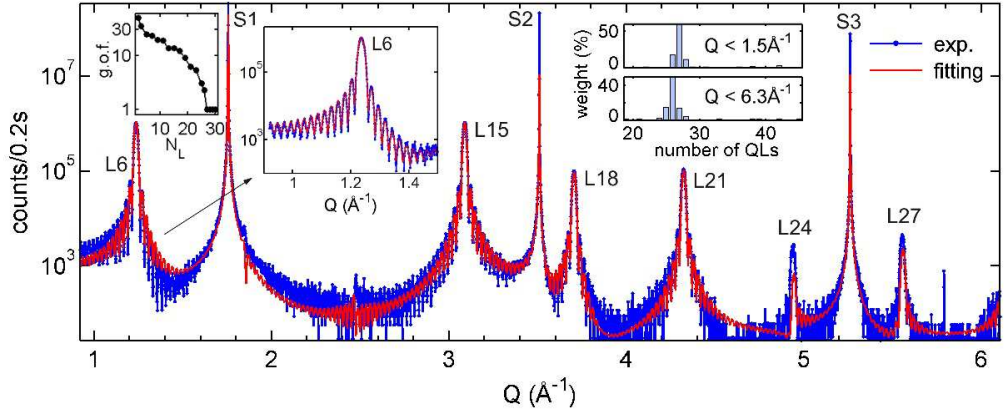}\\
  \caption{Experimental and fitted $Q$-scans along crystal truncation rods of 00L reflections in Bi$_2$Te$_3$ film on BaF$_2$ (111), sample S15057. Insets (from right to left): model structures in number of QLs covering the sample area (weight); best fit for the region $Q < 1.5\,\textrm{\AA}^{-1}$; and goodness of fit (g.o.f.) value as function of the number $N_L$ of layers in the model structures.}\label{fig:Qscan15057}
\end{figure*}

\begin{figure}
  \includegraphics[width=2.7in]{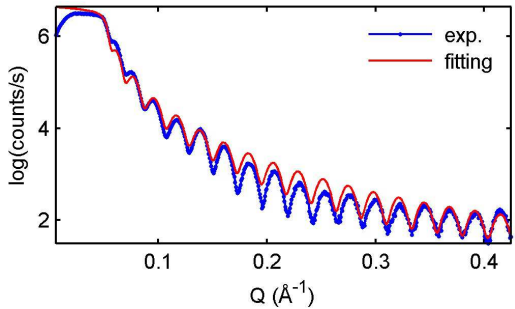}\\
  \caption{X-ray reflectometry in Bi$_2$Te$_3$ film on BaF$_2$ (111), sample S15057. Experimental curve measured in a PANalytical X'Pert MRD high-resolution X-ray diffractometer, Cu$K_\alpha$ radiation. Curve fitting using Eq.~(\ref{eq:R}) for a 26\,nm thick film, $\chi=1$, $d_i=0.57$\AA, and surface roughness of $2.2$\,\AA\, (see text).}
\label{fig:XRR15057}
\end{figure}

\begin{figure*}
  \includegraphics[width=5.4in]{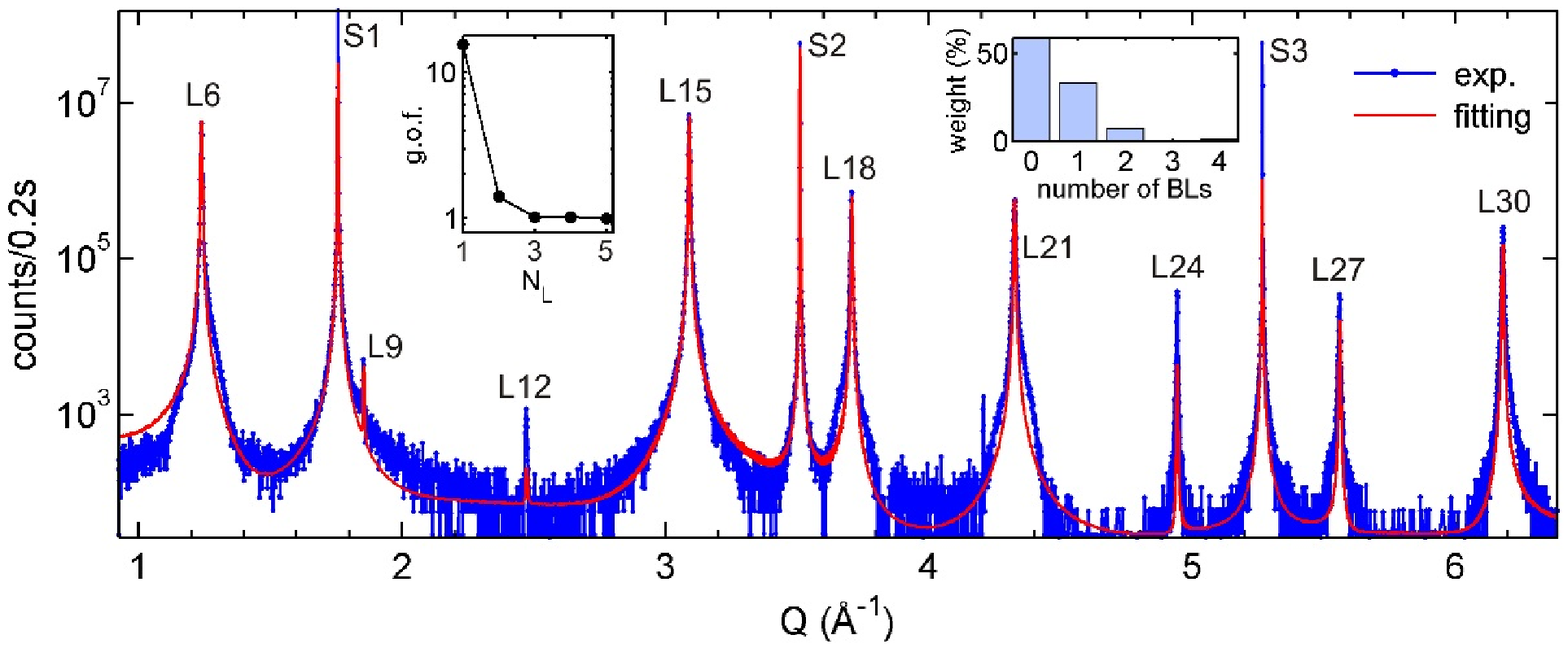}\\
  \caption{Experimental and fitted $Q$-scans along crystal truncation rods of 00L reflections in Bi$_2$Te$_3$ film on BaF$_2$ (111), sample S15065. Insets (from right to left): model structures for simulated data in number of BLs, M, on layers of composition (Bi$_2$)$_{\rm M}$(Bi$_2$Te$_3$)$_{165}$ covering the sample area (weight); and goodness of fit (g.o.f.) value as function of the number $N_L$ of layers in the model structures.}
\label{fig:Qscan15065}
\end{figure*}

Experimental and simulated data from three samples with Bi$_2$Te$_3$ epitaxial films are shown in Figs.~\ref{fig:Qscan15058}-\ref{fig:Qscan15065}. Nominally, these samples labeled S15058, S15057, and S15065 have films of thicknesses corresponding to the deposition time of 10QLs, 25QLs, and 165QLs, respectively. In all long-range $Q$-scans, the simulated data, i.e. the reflectivity curves as obtained from Eq.~\ref{eq:R}, are normalized to the experimental data by using as reference only the maximum intensity of peak L15. Refinement of model structures is based on a root-mean-square error of the log transformed data, $\zeta=(100/N_p)\sqrt{\sum_n[\log(I_n)-\log(R_n)]^2}$ where $I_n$ and $R_n$ are the $n$th experimental and simulated data points, respectively, and $N_p$ is the number of data points. For sake of visualization, a goodness of fit (g.o.f.) value given by $100(\zeta-\zeta_{\rm ref})^{1/2}$ has also been used here where $\zeta_{\rm ref}$ is an arbitrary value chosen to enhance the graphical perception of improving the g.o.f. value.

For the thinnest film, Fig.~\ref{fig:Qscan15058}, model structures with a single layer of uniform thickness provide fringes much more pronounced than observed in the experimental data; it can be seen by comparing the Q-scan in Fig.~\ref{fig:Qscan15058} with the simulated one shown in Fig. 4b. A drastic improve in fit quality, central inset in Fig.~\ref{fig:Qscan15058}, is obtained by adding simulated intensities from a number $N_L$ of layers of different thicknesses given in terms of an integer number of QLs. A model structure with one layer ($N_L$ = 1) covering 100\% of the sample area has a best fit for 7 QLs, interface distance $d_i = 1.88$\,\AA, and coherence $\chi = 1$. The improvement in fit quality goes until model structures containing twelve layers ($N_L = 12$) with thicknesses varying from 5 to 16 QLs. Considering more layers, layers of other thicknesses, or layers with other surface terminations than a complete QL have provided no gain in fit quality. A genetic algorithm is used to optimize the weight of each layer's contribution to the diffracted intensity, see \S\ref{ap:dfitt}.

By interpreting the values of the weights as fraction of surface area covered by the corresponding layer, we arrive that most of the sample area is covered by 7 QLs (72\%), followed by 6 QLs (11\%), 8 QLs (4\%), and 5 QLs (3\%). The remaining 10\% of area has minor contributions of a variety of layers, see right inset in Fig.~\ref{fig:Qscan15058} ($Q < 6.3\,\textrm{\AA}^{-1}$). However, a different set of weights is obtained when improving the fit quality only in the region $Q < 1.5\,\textrm{\AA}^{-1}$, see right inset in Fig.~\ref{fig:Qscan15058}. The best fit of this region, left inset in Fig.~\ref{fig:Qscan15058}, is achieved by using an interface distance $d_i = 0.84$\,\AA\, ($\chi = 1$), and a model structure that also has twelve layers ($N_L$ = 12) but with different weights: 7 QLs (30\%), 8 QLs (64\%), 10 QLs (2\%), 12 QLs (1\%), and 14 QLs (3\%).

A similar variation in model structures as a function of the analyzed scan range also occurs in sample S15057 with a thicker film, Fig.~\ref{fig:Qscan15057}. Improvement in fit quality goes until $N_L$ = 27, with thicknesses from 19 to 45 QLs. In the lower $Q$ region, $Q < 1.5\,\textrm{\AA}^{-1}$, the model with larger contribution of 27 QLs (60.3\%) and no layers thinner than 26 QLs (17.0\%) is responsible for the best fit of this region, see insets in Fig.~\ref{fig:Qscan15057}. On the other hand, when analyzing the entire scan range, major contributions are for 24 QLs (2.8\%), 25 QLs (14.4\%), 26 QLs (50.4\%), 27 QLs (13.2\%), and 28 QLs (4.3\%). Interface distance also varies, $d_i = 0.57$\,\AA\, ($Q < 1.5\,\textrm{\AA}^{-1}$) and $d_i = 1.57$\,\AA\, ($Q < 6.3\,\textrm{\AA}^{-1}$), $\chi = 1$ in both cases.

Model structures varying as a function of scan range reveal the limitation of Eqs.~\ref{eq:receqs} to simulate reflectivity curves in films with lateral structures or in-plane inhomogeneities. That is the same limitation faced when treating X-ray reflectometry by Fresnel reflection coefficients \cite{wor99,slm02}. For instance, reflectometry in the S15057 sample leads to a single layer model, as shown in Fig.~\ref{fig:XRR15057}. The strategy of adding intensities from layers of different thicknesses did not work at grazing incidence angles; it converges to a single layer of 26 QLs. That is consequence of interference between reflected amplitudes from distinct areas of the film. In laterally homogeneous films, the grazing incidence reflectivity curve can be well reproduced by attenuating the reflection coefficient with a Debye-Waller factor due to roughness. If that was the case, using $A=r_F\exp{(-Q^2u^2/2)}$ in Eq.~(\ref{eq:R}) would be enough to reproduce the reflectivity curve. But, the best fit achieved for a root mean square roughness of $u=2.2\pm0.2$\,\AA\, and $d_i = 0.57$\,\AA\, is still far from reproduce the experimental curve, which also compromise any attempt to physically interpret the small interface distances observed at low $Q$. This can be just a compensation for the interference effects between reflected amplitudes from adjacent film areas of different heights.

At higher incidence angles, the overlapping of reflected amplitudes from distinct areas are minimized and the strategy of adding intensities works better. Note that the thickness distributions are more pronounced when fitting $Q$-scans at higher angles, as observed for both films (insets of Figs.~\ref{fig:Qscan15058} and \ref{fig:Qscan15057}), and that at grazing incidence only one layer can be inferred, Fig.~\ref{fig:XRR15057}.   Therefore, investigating reflectivity curves at high angles is a procedure suitable to probe films with lateral structures as seems to be the case of Bi$_2$Te$_3$ films.

Besides of being laterally structured, other problem in analyzing bismuth telluride films is that thicker films are more likely to have BLs of bismuth. For sample S15065 with nominal thickness of 165 QLs, whose experimental $Q$-scan is presented in Fig.~\ref{fig:Qscan15065}, structure models with many layers of different thickness shows to be unfeasible. Minimization of the $\zeta$ function favors models with much thinner layers than 165 QLs in order to adjust the broad layer peaks that are seen in the experimental $Q$-scan. Although no split of the L21 peak is observed, a Te deficit can occur on small fractions over the sample area. To investigate this possibility, average curves from layers of composition (Bi$_2$)$_{\rm M}$(Bi$_2$Te$_3$)$_{165}$ with random distribution of BLs were simulated in the same manner of the scans in Fig.~\ref{fig:simQscanrand}. Then, the fit quality is improved by adding simulated intensities from a number $N_L$ of layers with different M values. With such models, the best fit curve shown in Fig.~\ref{fig:Qscan15065} is quickly achieved just after considering three layers ($N_L$ = 3) with compositions M = 0 (57.5\%), 1 (33.8\%), and 2 (8.7\%), insets in Fig.~\ref{fig:Qscan15065}. It implies in a film of mean composition Bi$_2$Te$_{2.991}$.

\begin{figure*}
  \includegraphics[width=5.4in]{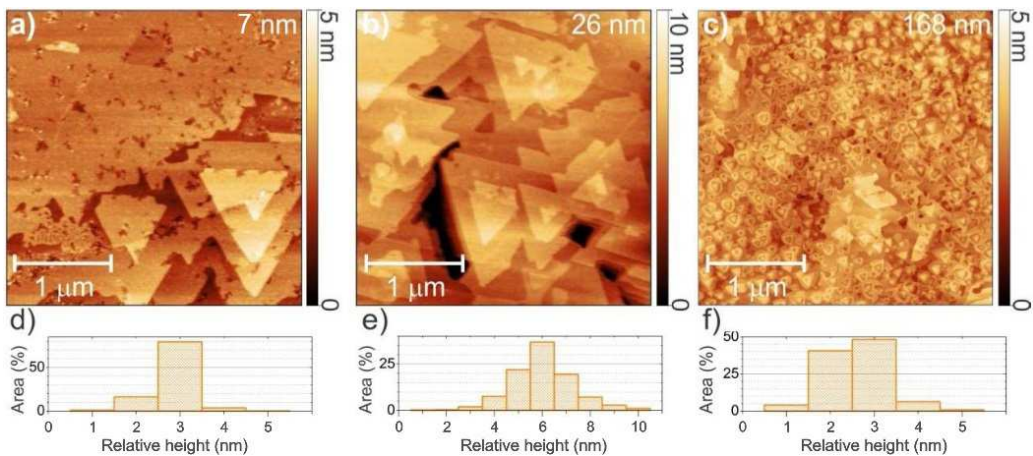}\\
  \caption{Atomic force microscopy images of Bi$_2$Te$_3$ epitaxial films, samples (a) S15058, (b) S15057, and (c) S15058. Most probable height according to X-ray diffraction simulation are indicated on each image. (d-f) Histograms of height distribution over the respective imaged area.}
\label{fig:AFM}
\end{figure*}

Surface morphology of pure Bi$_2$Te$_3$ films typically exhibit spiral-like triangular domains---reflecting the three-fold symmetry of hexagonal planes---with terraces steps of one complete QL, approximately 1\,nm in height \cite{kru11,ste14}. The typical domains are also seen in AFM images of our samples, Figs.~\ref{fig:AFM}a-c. However, these representative $3\times3\,\mu{\rm m}^2$ AFM images of the samples surfaces clearly show that the domains become smaller as the films thicknesses increase. In the thicker film, Fig.~\ref{fig:AFM}c, the triangular domains have dimensions that are about 10 times smaller than observed in the thinner films, Figs.~\ref{fig:AFM}a,b.

For sake of comparison with X-ray data, height distributions from AFM images are shown in Figs.~\ref{fig:AFM}d-f. Since the height scales of AFM measurements are relative, each scale has to be shifted by an offset before comparison with model structures from X-ray data. For both S15058 and S15057 samples, there are perfect match between the AFM and X-ray height distributions when considering the models obtained by fitting the large $Q$ ranges, $Q < 6.5\,\textrm{\AA}^{-1}$ insets in Figs.~\ref{fig:Qscan15058} and \ref{fig:Qscan15057}. On the other hand, in simulating X-ray data from the thicker film, sample S15065, it was necessary to consider only one layer (no thickness fluctuations). This result is in agreement with the AFM histogram in Fig.~\ref{fig:AFM}f where 90\% of the sample area has a thickness flutuation of only 1\,nm (1\,QL), which corresponding to 0.6\% of the nominal film thickness of 165\,QLs. Despite the fact that height distributions from AFM images are representative of small areas regarding the X-ray footprint at the sample surfaces, these results corroborate very well with the ones obtained from the X-ray diffraction simulation. It validates our diffraction models of adding reflectivities instead of reflected amplitudes from layers of different thicknesses, at least at high $Q$ values.

There are however two intriguing facts in the X-ray and AFM data of the thicker film. Bragg peaks of the film in Fig.~\ref{fig:Qscan15065}, mainly peaks L6, L15, and L18, present asymmetric broadening at their shoulders that can not be addressed by our model structures built up of layers either with different thicknesses or compositions. It is possible that an odd distribution of bismuth BLs could produce such broadening, but check this possibility is far from what can be done by model structures with a few tens of layers; too many layers produce unreliable models by fitting algorithms. The other possibility is that thicker films have developed a lateral structure of defects during growth, in agreement with the AFM images.

\begin{figure}
  \includegraphics[width=2.7in]{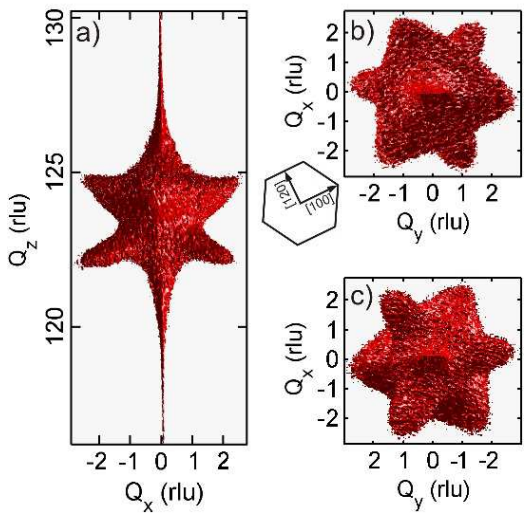}\\
  \caption{Isointensity surface in reciprocal space of peak L6 in Bi$_2$Te$_3$ film on BaF$_2$ (111), sample S15065. (a) Side view, (b) top view, and (c) bottom view of the reciprocal lattice node. $Q_z$ is along the crystal truncation rod. $Q_x$ and $Q_y$ are along in-plane directions parallel and perpendicular to the incidence plane, respectively. Inset: crystallographic directions with respect to the view in (b). Intensity level of 1000 cps computed from 150 images acquired by rocking the sample in steps of 0.02$^{\circ}$. rlu=0.01\,\AA$^{-1}$.}
\label{fig:rlp}
\end{figure}

The three-dimensional growth of triangular domains observed in the AFM images evolves from domains with large plateaus in thin films, Figs.~\ref{fig:AFM}a and \ref{fig:AFM}b, to very small ones in thicker films, Fig.~\ref{fig:AFM}c, indicating that new layers start to grow before one is completed. Such non layer-by-layer growth mechanism favors formation of stacking faults at domain boundaries. Because our X-ray data collection was carried out with an area detector, it is quite easy to acquire the diffracted intensities over the detector area as a function of the rocking curve of the sample. This procedure allows to reconstruct the three-dimensional aspect of the reciprocal lattice node \cite{som09}, as shown in Fig.~\ref{fig:rlp} ---a standard reciprocal space map that integrates the axial intensity distribution, along $Q_y$ in the figure, would be a two-dimensional projection of the image in Fig.~\ref{fig:rlp}a---. Quite surprisingly, the 3D shape of the node is well correlated with the triangular shape of the domains observed in the AFM images. In-plane crystallographic directions in Fig.~\ref{fig:rlp} (inset) were determined by measuring the $1\,\bar{1}\,20$ Bi$_2$Te$_3$ asymmetric reflection. Two distinct features dominate the observed node shape, an upside down tetrahedron whose base is better seen in the top view, Fig.~\ref{fig:rlp}b, and three tips emerging from the inclined faces of the tetrahedron that are better seen in the bottom view, Fig.~\ref{fig:rlp}c. The tetrahedral shape seems to be caused by truncation (Fourier transform) of diffracting domains with pyramidal form, while the tips resemble the intensity streaks caused by stacking faults on inclined planes \cite{pk06,sb15}. In this case, the upper part of the tips/streaks would be inside the tetrahedron. Each tip and its opposite tetrahedron vertex are approximately oriented at 60$^{\circ}$ from the vertical direction in Fig.~\ref{fig:rlp}a, closely aligned along the normal direction of Bragg planes $(1\,\bar{1}\,5)$, $(0\,1\,5)$, and $(\bar{1}\,0\,5)$, which are planes of a same family according to the hexagonal indexing notation, $(1\,\bar{1}\,0\,5)$, $(0\,1\,\bar{1}\,5)$, and $(\bar{1}\,0\,1\,5)$ planes, respectively. These planes stand for the strongest reflections in hexagonal Bi$_2$Te$_3$ crystal.

It is also possible that the tips are related to twinned domains that have been also reported in this epitaxial system \cite{for16}. But, only if these domains differ in shape regarding of the normal domains, such as upper side down tetrahedra twisted by $60^\circ$ around the growth direction. In thin films (samples S15058 and S15057) or in thick films with large domains \cite{cif16} there is no occurrence of lateral features in the reciprocal lattice nodes. Further investigation would be necessary before we draw a direct correlation between the detailed shapes of reciprocal lattice nodes and the nanostructure of domains in the films. But in regard to this work, the fact that the analyzed node has no symmetrical intensity distribution along the specular direction provide an explanation for the discrepancy between  experimental and simulated scans of the thicker film in Fig.~\ref{fig:Qscan15065}.

\section{Conclusions}

Simulation of wide X-ray diffraction along crystal truncation rods in specular geometry has been possible by accounting for refraction, absorption, and rescattering events between any atomic layers. As a consequence, very weak and very strong reflections are treated likewise within this approach, which is suitable to be implemented in auto-fitting algorithms for refinement of the model structures. Here, the auto-fitting procedure leads to film structures with surface area covered by many layers either of different thicknesses or compositions. In thin films, all layers start and end with complete quintuple layers. In the thick film, occurrence of nanometer sized domains are responsible for broadening of reciprocal lattice nodes in inclined directions even in the absence of mosaicity in the film. Reciprocal space mapping with area detector provides rich information about the nanostructure of domains in the films, opening new opportunities for better understanding of the growth dynamics in epitaxic layers.

\begin{acknowledgments}
The authors acknowledge CNPq (Grant Nos. 142191/2014-0, 302134/2014-0, 307933/2013-0, and 306982/2012-9) and FAPESP (Grant Nos. 2014/04150-0 and 2016/11812-4) for financial support.
\end{acknowledgments}

\appendix
\section{Supplemental Information}

\subsection{Atomic inter-layer distances}\label{ap:ipdist}

In epitaxic $({\rm Bi}_2)_{\rm M}({\rm Bi}_2{\rm Te}_3)_{\rm N}$ films on BaF$_2$ (111), the mean atomic inter-layer distance $\langle d \rangle=d_0+b_0\delta$ has been reported to vary linearly with composition $\delta=3{\rm M}/({\rm N}+{\rm M})$ \cite{ste14}, as well as each one of the inter-layer distances $d_n=d_{n,0}+b_n\delta$ in Fig.~\ref{fig:nQLSL}. Since $d_1=d_4$, $d_2=d_3$, and $$\langle d \rangle = \dfrac{ND_5+M(D_7-D_5)}{5N+2M}$$ where $D_5=2(d_1+d_2)+d_5$ and $D_7=2(d_1+d_2+d_6)+d_7$ are the thicknesses of the nQL and SL sets of MLs, a few constraints can be applied for refining experimental values of the inter-layer distances. These constraints are $2d_{1,0}+2d_{2,0}+d_{5,0} = 5d_0$, $2b_6+b_7-2(b_1+b_2+b_5) = -3b_0$, and $2d_{6,0}+d_{7,0}-d_{5,0}+3[2(b_1+b_2)+b_5]=2d_0+15b_0$. Due to the very small mismatch of in-plane lattice parameters, we use near bulk values for $d_{1,0}=1.746$\,\AA, $d_{2,0}=2.035$\,\AA, and $d_{5,0}=2.613$\,\AA\, \cite{sn63,kuz15}. It leads to $d_0=2.035$\,\AA, and to a good match for the experimental positions of L peaks in films with no Te deficit, Figs.~\ref{fig:Qscan15058} and \ref{fig:Qscan15057}. By applying the above constraints to the observed linear behavior of $d_n$ where $b_2\simeq0$, $b_1\simeq-b_5$, and $b_7\simeq-b_6$ we end up with $b_6=-3b_0$, and hence
\begin{equation}
    \left[
      \begin{array}{c}
        \langle d \rangle \\
        d_1 \\
        d_2 \\
        d_5 \\
        d_6 \\
        d_7 \\
      \end{array}
    \right]=
    \left[
      \begin{array}{c}
        2.035 \\
        1.746 \\
        2.035 \\
        2.613 \\
        2.178 \\
        2.027 \\
      \end{array}
    \right]+
    \left[
      \begin{array}{r}
        -0.025 \\
        -0.025 \\
        0 \\
        0.025 \\
        0.075 \\
        -0.075 \\
      \end{array}
    \right]\delta\,\textrm{\AA} \nonumber
\end{equation}
is one solution for $b_0=-0.025$\,\AA\, when also using that $b_5\simeq b_6/3$ and $d_{6,0}/(2d_{6,0}+d_{7,0})=0.3412$. Within experimental accuracy of reported values, other solutions are possible including one in which $b_1=b_5=0$ as used in this work.

\subsection{Rescattering effects}\label{ap:rescatt}

\begin{figure}
\includegraphics[width=2.7in]{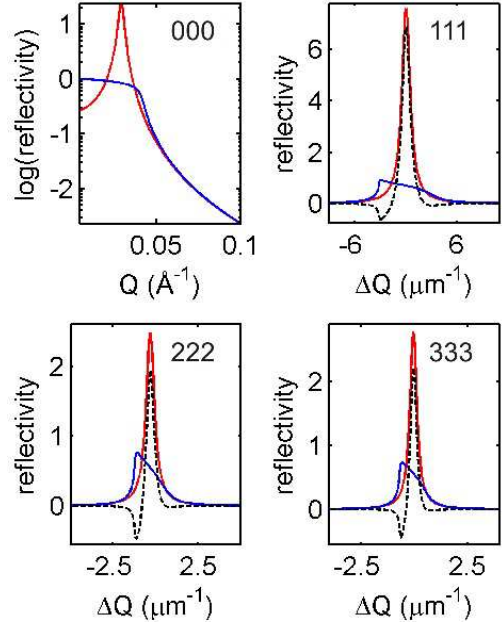}
\caption{Comparison of X-ray reflectivity curves in BaF$_2$ (111) substrate when accounting for the rescattering term $\bar{r}_A r_B$ in Eqs.~(\ref{eq:receqs}) (blue lines), or when neglecting this term, $\bar{r}_A r_B=0$ (red lines). Curves differences are also shown (dashed lines).}
\label{fig:BaF2peaksLLL}
\end{figure}

By playing with the $\bar{r}_A r_B$ term in Eqs.~(\ref{eq:receqs}), we can identify regions of the specular reflectivity curve that are most susceptible to effects of photon rescattering. When using this term, we have the exact solution of the dynamical theory in the entire reflectivity curve since the grazing incidence region until the intrinsic profiles of the Bragg peaks at high angles. On the other hand, when setting this term null, $\bar{r}_A r_B=0$, only photon rescattering is been neglected; absorption and refraction are still accounted for in Eqs.~(\ref{eq:receqs}). For the BaF$_2$ (111) substrate, rescattering effects are critical at the total refraction region, in a range of about $Q<0.1\,\textrm{\AA}^{-1}$, and within the FWHM of Bragg peaks, as shown in Fig.~\ref{fig:BaF2peaksLLL}. In the data fitting procedure described below, \S\ref{ap:dfitt}, the rescattering term have to be considered only when fitting the total refraction region, e.g. Fig.~\ref{fig:XRR15057}. This term has no effect on model structures obtained by fitting the reflectivity curves around Bragg peaks.

\subsection{Scattering by monoatomic layers}\label{ap:scattmls}

Non-resonant and resonant terms of the atomic scattering factors in $f_a(Q,E)=f_a^0(Q)+f_a^{\prime}(E)+if_a^{\prime\prime}(E)$ are from tabulated values \cite{ep06}, although they were effectively calculated by routines asfQ.m and fpfpp.m that can be found in open codes at the internet \cite{slm16}. In both substrate and epilayer materials, each ML plane contain a single element so that the number of atoms per area unit are $\eta_{\rm Ba}$ = $\eta_{\rm F}$ = 1/16.6445\,atoms/\AA$^2$ and $\eta_{\rm Bi}$ = $\eta_{\rm Te}$ = 1/16.6294\,atoms/\AA$^2$. In the Debye-Waller factors multiplying $f_a(Q,E)$, the root-mean-square displacement value of 0.1\,\AA\, was used for both Ba and F atoms, while for Bi and Te atoms it was 0.19\,\AA.

\subsection{Data fitting}\label{ap:dfitt}

X-ray reflectivity in epitaxial Bi$_2$Te$_3$ films with no Te deficit were computed as
\begin{equation}\label{eq:wR}
    R({\bm p},Q) = \frac{\sum_{j=1}^{N_L} w_j R_j({\bm p},Q)}{\sum_{j=1}^{N_L} w_j}
\end{equation}
where $R_j({\bm p},Q)$ is the X-ray reflectivity given by Eq.~(\ref{eq:R}) for the $j$th layer in the model structure, being that each layer $j$ has an integer number of complete QLs. ${\bm p}$ stands for the set of input parameters that are common to all layers: X-ray energy $E$, substrate lattice parameter $a_S$, interface distance $d_i$ and coherence $\chi$, $Q$-scan offset, and surface termination. These parameters have been either provided or adjusted manually. A differential evolution algorithm already optimized for fitting X-ray data \cite{wor99} have been used to weigh the contribution $w_j$ of each layer. The evolution of the curve fitting is guided by minimization of the $\zeta$ function, previously defined as the root-mean-square error of the log transformed data.

For Bi$_2$Te$_{3-\delta}$ films with deficit $\delta$ of Te, occurrence of bismuth bilayers (BLs) is assumed to follow a log-normal distribution, i.e. the integer number $n$ of QLs in between any two consecutive BLs is given by the probability density function
\begin{equation}\label{eq:lognorm}
    P(n) = \frac{1}{n\,\sigma\,\sqrt{2\pi}}\,\exp{\left[-\frac{(\ln n -\ln b)^2}{2\sigma^2} \right]}
\end{equation}
where $b = n_0\exp(\sigma^2)$, $n_0$ is the most probable number of adjacent QLs, and $\sigma$ is the standard deviation value in logarithmic scale. A summary of the $n_0$ and $\sigma$ values used in this work is given in Table~\ref{tab:n0sigma}.

\begin{table}
\begin{center}
\begin{tabular}{cccccccc}
\hline\hline
Fig. & \ref{fig:simQscanrand}a & \ref{fig:simQscanrand}b & \ref{fig:simQscanrand}c & \ref{fig:simQscanrand}d & \ref{fig:simQscanrand}e & \ref{fig:QLdistrib}a & \ref{fig:Qscan15065} \\
\hline
$n_0$ & 84 & 28 & 15 & 8 & 8 & 8 & 82, 55, 41, and 33 \\
$\sigma$ & 0.6 & 0.6 & 0.6 & 0.6 & 0.05 & 0.6 & 0.6 \\
\hline\hline
\end{tabular}
\caption{Values of $n_0$ and $\sigma$ used is this work.}
\label{tab:n0sigma}
\end{center}
\end{table}

%\bibliography{bitemorelhao-novo}

\end{document}